\documentstyle[twocolumn,aps]{revtex}
\begin{document}
\input psfig
\draft

%
\twocolumn[\hsize\textwidth\columnwidth\hsize\csname@twocolumnfalse\endcsname
%
%

\title{Effect of Non-Magnetic Impurities (Zn, Li) 
\\ in a Hole Doped 
Spin-Fermion Model for Cuprates}

\author{Charles Buhler$^1$, Seiji Yunoki$^2$ and Adriana Moreo$^1$}

\address{$^1$Department of Physics, National High Magnetic Field Lab and
MARTECH,\\ Florida State University, Tallahassee, FL 32306, USA}

\address{$^2$Materials Science Center, \\ University of Groningen,
Nijenborgh 4, 9747 AG Groningen, The Netherlands}

\date{\today}
\maketitle

\begin{abstract}

The effect of adding non-magnetic impurities (NMI), such as Zn or Li, to 
high-Tc cuprates is studied applying Monte Carlo techniques to a spin-fermion 
model. It is observed that adding Li is qualitatively similar to
doping with equal percentages of Sr and Zn. The mobile holes (MH) are trapped 
by the NMI and the system remains insulating and
commensurate with antiferromagnetic (AF) correlations. This
behavior persists in the region ${\rm \%NMI>\%MH}$. On the other hand,
when ${\rm \%NMI < \%MH}$ 
magnetic and charge 
incommensurabilities are observed. The vertical or
horizontal hole-rich stripes, 
present when \% NMI=0 upon hole doping, are pinned by the 
NMI and tend to become diagonal, surrounding finite AF domains. 
The \%MH-\%NMI 
plane is investigated. Good agreement
with experimental results is found in the small portion of this 
diagram where experimental data are available.
Predictions about the expected behavior in the 
remaining regions are made.

\end{abstract}

\vskip2pc]
\narrowtext

Neutron scattering studies of the high-Tc cuprates have established that 
spin incommensurability (IC) appears in these compounds upon hole 
doping.\cite{Cheong} There is mounting evidence indicating that this IC
is due to the formation of charge stripes.\cite{Tran}
One way of understanding the relationship between magnetic and charge 
fluctuations, as well as their impact in other properties of these materials, 
is by introducing NMI in the Cu-O planes. Some of 
these experimental results are puzzling. It has been observed
that replacing ${\rm Cu^{2+}}$ by ${\rm Zn^{2+}}$ 
depletes antiferromagnetism without causing spin IC even at 25\%
doping.\cite{Chakra} It is estimated that an impurity content as large as 
30-40\% will be required to destroy AF order in this case.
\cite{Hucker,Fisk} While the addition of MH 
(Sr in ${\rm La_{2-x}Sr_xCuO_4}$ (LSCO))
to lightly Zn-doped materials 
produces spin IC similar to the one observed in LSCO,
\cite{Hirota} doping with Li, which introduces equal numbers of localized 
impurities and mobile holes does not induce spin IC 
at 10\% Li doping of ${\rm La_2CuO_4}$
(LCO).\cite{Bao}

The goal of this paper is to gain theoretical understanding on the effect of
NMI doping on cuprates.
Numerical studies of Hubbard and t-J models,
the Hamiltonians traditionally used to study the physics of the cuprates, 
are very difficult at low temperatures. A simpler 
spin-fermion model has been used successfully to understand magnetic 
and charge properties in these materials, showing that spin IC and charge
stripe formation are related.\cite{Charlie}
The spin-fermion model is constructed as an interacting system of
electrons and spins, crudely mimicking phenomenologically the
coexistence of charge and spin degrees of freedom in 
the cuprates.\cite{Pines,Schrieffer,Fedro}. Its Hamiltonian is given by
$$
{\rm H=
-t{ \sum_{\langle {\bf ij} \rangle\alpha}(c^{\dagger}_{{\bf i}\alpha}
c_{{\bf j}\alpha}+h.c.)}}
+{\rm J
\sum_{{\bf i}}
{\bf{s}}_{\bf i}\cdot{\bf{S}}_{\bf i}
+J'\sum_{\langle {\bf ij} \rangle}{\bf{S}}_{\bf i} \cdot{\bf{S}}_{\bf j}},
\eqno(1)
$$
\noindent where ${\rm c^{\dagger}_{{\bf i}\alpha} }$ creates an electron
at site ${\bf i}=({\rm i_x,i_y})$ with spin projection $\alpha$,  
${\bf s_i}$=$\rm \sum_{\alpha\beta} 
c^{\dagger}_{{\bf i}\alpha}{\bf{\sigma}}_{\alpha\beta}c_{{\bf
i}\beta}$ is the spin of the mobile electron, the  Pauli
matrices are denoted by ${\bf{\sigma}}$,
${\bf{S}_i}$ is the localized
spin at site ${\bf i}$,
${ \langle {\bf ij} \rangle }$ denotes nearest-neighbor (NN)
lattice sites,
${\rm t}$ is the NN-hopping amplitude for the electrons,
${\rm J>0}$ is an AF coupling between the spins of
the mobile and localized degrees of freedom,
and ${\rm J'>0}$ is a direct AF coupling
between the localized spins.
The density $\rm \langle n \rangle$=$\rm 1-x$ of 
itinerant electrons is controlled by a chemical potential $\mu$. 
Hereafter ${\rm t=1}$ will be used as the unit of energy. 
>From 
previous phenomenological analysis the coupling ${\rm J}$ 
is expected to be larger than t, while the Heisenberg coupling
${\rm J'}$ is expected to be smaller.\cite{Schrieffer,Fedro,Charlie} 
The value of ${\rm J}$ will be here fixed to 2 and
the coupling ${\rm J'=0.5}$, as in Ref.\cite{Charlie}.
To simplify the numerical calculations, avoiding the sign problem, 
localized spins are assumed to be classical (with $\rm |S_{\bf i}|$=1).
This approximation is not as drastic as it appears and it was discussed in 
detail in Ref. \cite{Charlie}. The model will be studied using a
Monte Carlo (MC) method, details of which can be found in Ref.~\cite{yuno}.
Periodic boundary conditions (PBC) are mainly 
used but it has been checked that 
the results presented here are independent of the boundary conditions.

Doping with Zn will be simulated by randomly removing localized spins. 
This is achieved by turning $J=0$ at the impurity site and $J'=0$ along  
the four links that connect the impurity to neighboring sites.
Li doping has to be treated differently. Since it is in an 
oxidation state ${\rm Li^+}$ (while Cu and Zn are 2+), the ion effectively 
acts as a negative charge 
center. To simulate this effect the hopping $t$ that connects the impurity
to neighboring sites should be reduced.
In addition, the Li ion is closed shell and thus, non-magnetic. 
This
feature is simulated as in the case of Zn, by randomly removing localized
spin degrees of freedom. An additional effect of Li doping 
is the introduction of one MH, thus the chemical potential has to
be tuned such that the percentage of mobile holes equals the percentage of
localized impurities.

To study the magnetic properties of the system the spin-spin
correlation functions among the classical spins defined as
$\rm \omega({\bf r})={1\over{N}}\sum_{\bf i}\langle{\bf{S}}_{\bf
i}\cdot{\bf{S}}_{\bf i+r}\rangle$
(where N is the number of sites) was measured. Its
Fourier transform provides the spin structure
factor $S({\bf q})$.
The momentum $\rm q_\gamma$ takes the values ${\rm 2 \pi
n/L_\gamma}$, with
${\rm n}$ running from 0 to ${\rm L_\gamma-1}$, and ${\rm L_\gamma}$ denoting 
the number of sites
along the $\gamma$=x or y direction. 
\begin{figure}[thbp]
\centerline{\psfig{figure=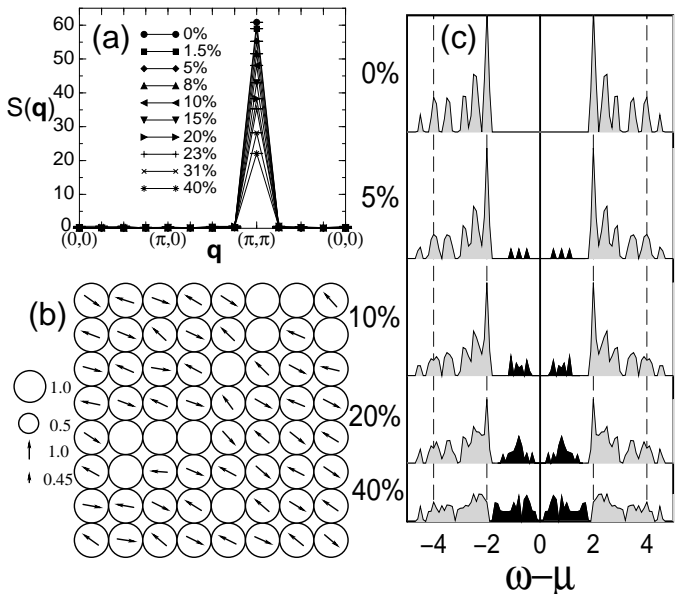,height=8cm}}
\caption{(a) Structure factor $S({\bf q})$ for the localized spins versus
momentum, for J=2, $\rm J'$=0.05 and T=0.01 on an 8$\times$8
lattice and several Zn concentrations at $\rm \langle n \rangle$=1 along
$(0,0)-(0,\pi)-(\pi,\pi)-(0,0)$; (b) spin and
charge distribution for a typical MC 
snapshot for 15\% Zn and the
same parameters as in (a). The arrow lengths are proportional to the 
projection of the localized spin
on the (X-Z) plane, and
the radius of the circles is proportional to the local electronic density
$n({\bf i})$, according to the scale
shown. PBC are used. (c) Density of states for different percentages of
Zn doping at $\langle n \rangle=1$ 
with the same parameters as in (a). The dark areas 
indicate states with very little dispersion}
\end{figure}
In Fig.1-a the structure factor for different concentrations of localized 
NMI, like Zn, is shown for
an $8 \times 8$ lattice at temperature $T=0.01t$ and at 
$\rm \langle n \rangle$=1 since doping with Zn
does not modify the density of itinerant electrons. It can be seen that 
the peak remains at $(\pi,\pi)$ for all the studied NMI concentrations. The
only effect of adding NMI is to reduce the intensity of the structure factor
since spins are being removed.
It is clear that random impurities do not destroy the AF 
order as can be observed from a snapshot at 15 \% doping for the spins on the 
plane X-Z 
(Fig.1-b). The snapshot also shows a uniform charge distribution.
This is in complete agreement with experimental data for Zn
doping in LCO.\cite{Hirota} Our simulations can produce dynamical 
information directly in real-frequency
without the need of carrying out (uncontrolled) analytic extrapolations from
the imaginary axis. This allows us to study the behavior of the 
density of states (DOS). In Fig.1-c the DOS as a function of 
\%NMI doping is presented. The addition of localized impurities creates 
states inside the gap which are indicated by a black filling in the figure. 
Analyzing the spectral functions $A({\bf q},\omega)$, it was 
observed that these
states have very small dispersion which added to the absence of spectral 
weight at the chemical potential indicates that 
the system remains an insulator, in agreement with 
experiments.\cite{Chakra}
\begin{figure}[thbp]
\centerline{\psfig{figure=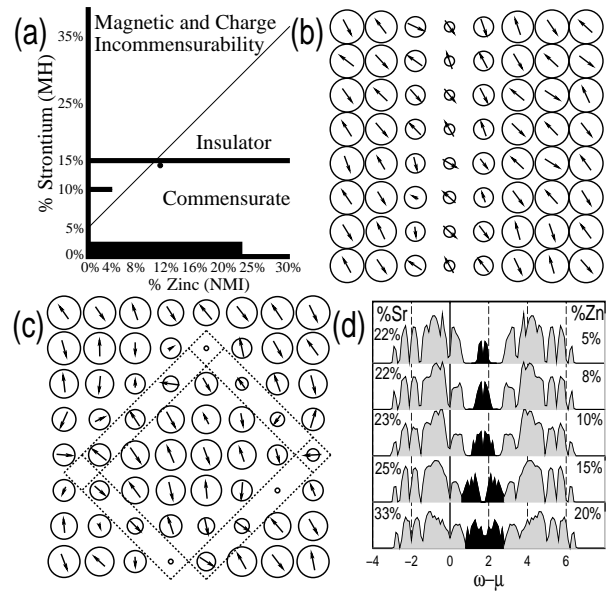,height=8cm}}
\caption{(a) Schematic magnetic and charge phase diagram
in the plane ${\rm \%Sr-\%Zn}$. 
The thick lines and the dark areas indicate the regions that have been 
studied experimentally; 
(b) Spin and
charge distribution for a MC snapshot on an 8$\times$8 cluster 
for J=2, $\rm J'$=0.05, T=0.01, at $\langle {\rm n} \rangle$$\approx$
0.85 and \% Zn=0. The spins are shown on the X-Y plane; (c) same as (b) but 
with \% Zn=5; (d) DOS for different cases with \%Sr$>$\%Zn. The notation is
as in Fig.1. PBC are used}
\end{figure}

The next step in our investigation is to study magnetic and charge properties 
varying both the mobile hole (Sr) and localized NMI (Zn) 
densities. In this case only a small fraction of the \%Sr-\%Zn
plane has been explored experimentally. Studies were performed for $0\leq
\%{\rm Zn} \leq 25$ and $0 \leq {\rm \%Sr} \leq 3$ \cite{Hucker}, 
${\rm 0 \leq \% Zn \leq 75}$ 
and $\% {\rm Sr}=15$ \cite{Hilscher}, 12\% Zn and 14\%
Sr \cite{Hirota}, $0\leq\%{\rm Zn} \leq 12$ and 15\% Sr \cite{Kar}, 
$0\leq\%{\rm Zn} \leq 4$ and 10\% Sr \cite{Fuku}, $0\leq\%{\rm Zn} \leq 40$ and
0\% Sr \cite{Chakra}. Numerous experiments have also 
been performed on LSCO (without
Zn). For ${\rm 0 \leq \% Sr \leq 25}$ neutron 
scattering experiments are presented
in Ref.\cite{yamada}. Zn-doped ${\rm YBa_2Cu_3O_{7-\delta}}$ (YBCO) 
has also been studied with $0\leq\%{\rm Zn} 
\leq 4$ and the O content varying between 6.6 and 7, which roughly 
corresponds to a percentage of MH ranging between 9-15\%.
\cite{YBC,Fuku}. These regions are shown in Fig.2-a.
In the same figure the magnetic and charge 
properties of the model Eq.(1) in the plane ${\rm \% Sr-\% Zn}$ 
obtained from the present study are presented. 
In the absence of NMI, the addition of 
MH induces stripe formation, as well as spin IC. The stripes are either
horizontal or vertical\cite{Charlie} and a snapshot for 15\% MH 
and no localized impurities is shown 
in Fig.2-b. As it can be seen in the figure, the electronic density
is depressed near the stripe indicating that perpendicular fluctuations
are present, i.e., there is short range movement of holes in the direction
perpendicular to the stripe.
 
When localized NMI are added
they tend to trap holes. This reduces the effective hopping towards the
impurity's 
nearest neighbor sites and diagonal stripes anchored by the impurities 
tend to develop as boundaries of AF domains. This effect is similar to the 
formation of diagonal stripes due to pinning experimentaly observed. 
\cite{Tran2} 
An example of this behavior is shown
in Fig.2-c where 3 NMI (i.e. 5\% Zn) have been added to the system with
$\langle n \rangle\approx 0.85$. 
A $\pi-$shift also occurs between the AF domains 
separated by the diagonal stripes. For some distributions of Zn impurities
both diagonal and non-diagonal stripes, pinned by the impurities, coexist. 

\begin{figure}[thbp]
\centerline{\psfig{figure=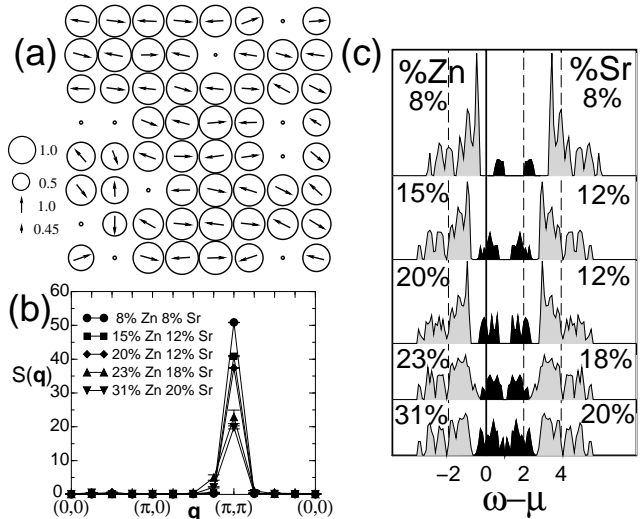,height=7cm}}
\caption{(a) Snapshot for \%Zn=\%Sr=15 in the plane X-Z; 
(b) S(q) for different points with
${\rm \%Zn\geq \%Sr}$; (c) Density of states for the parameters shown in (b).
The parameters and notation are as in Fig.1}
\end{figure}

It has also been observed that while a localized impurity does not disturb the 
AF background, when the number of mobile holes is larger than
the number of localized impurities a $\pi-$shift occurs in 
the x and y directions across the impurity which eventually will give rise 
to spin IC (see Fig.2-c)\cite{com,e2}. According to this result, spin IC 
should be detected
in the region ${\rm \% Sr>\% Zn}$, more specifically, 
when the percentage of Sr 
is larger than that of Zn
by approximately 5\%
according to experimental results for LSCO. This is in agreement with our 
observations if it is taken into account that due to the finite size of our
system, spin IC is first detected when the peak in the structure factor is at
$(\pi,3\pi/4)$ and symmetrical points. Thus, it is here predicted 
that spin IC should be
experimentally observed in samples with ${\rm \% Sr>\% Zn}$. Notice 
that, due to the fact that the stripe structure is modified by NMI doping
this may not be the same kind of IC observed in LSCO. In fact, it is
possible that besides the magnetic peaks at $(\pi-\delta,\pi)$ and 
symmetrical points, less intense peaks may appear at 
$(\pi-\delta,\pi-\delta)$. More
generally, as the percentage of NMI increases,
IC peaks maybe observed in all directions in momentum space.
The DOS 
indicates that in this situation the system has metallic characteristics 
although the chemical potential is in a pseudogap as shown in Fig.2-d. This 
feature is also in agreement with experiments.\cite{endoh} A study of the
spectral function shows that the states indicated with black, have very little
dispersion. From the figure it can be seen that some mobile holes are 
trapped in these
states and become almost localized but the extra mobile holes go into the 
dispersive states that form the pseudogap.

\begin{figure}[thbp]
\centerline{\psfig{figure=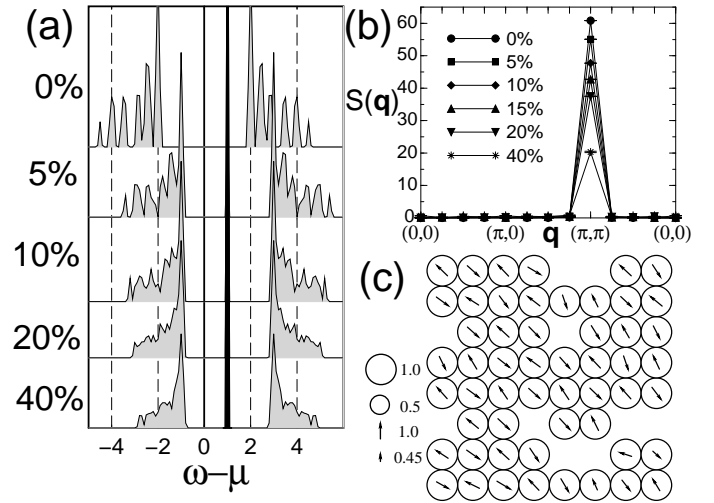,height=7cm}}
\caption{(a) Density of states for different percentages of Li doping
on an 8$\times$8 cluster 
for J=2, $\rm J'$=0.05, T=0.01;(b) Structure factor for the localized
spins as a function of the momentum. The parameters are the same as in (a); 
(c) snapshot for \%Li=15 (same parameters as in (a) and same notation as in
Fig.1.)}
\end{figure}

When ${\rm \%MH\leq \% NMI}$, the spinless 
centers tend to localize the mobile holes
as it is shown in Fig.3-a for ${\rm \%Zn=\%Sr=15}$. 
Though the hopping towards the
impurity sites is $t=1$, the electrons prefer to stay away from that site,
since the coupling J is zero and the observed electronic density in them
is just $\langle n_{\bf i}\rangle \approx 0.03$, which indicates 
that the holes are very localized at the impurity sites. 
Notice that the hole localization, which is observed experimentally
\cite{Hilscher,Hucker}, appears naturally in our calculations, while it has to 
be assumed in other theoretical approaches.\cite{Neto}
The system is magnetically commensurate
and the structure factor peaks at $(\pi,\pi)$ as shown in Fig.3-b for
different values of ${\rm \%Zn\geq\%Sr}$. 
In these circumstances the system is an
insulator as it can be seen in Fig.3-c where the DOS is displayed. When 
\%Zn=\%Sr the chemical potential lies in a gap separating dispersive
from non-dispersive states. When ${\rm \%Zn>\%Sr}$ the chemical potential
is located at the non-dispersive states. In both cases the
system is an insulator. 
As mentioned before, very few 
experiments where performed in this region of the plane. However, our 
results are in agreement with the available data since insulating
behavior was found for ${\rm \%Sr=15}$ and
${\rm \%Zn\geq 15}$ 
Ref.\cite{Hilscher} and ${\rm \%Sr=3}$ and ${\rm \%Zn\geq 3}$ 
Ref.\cite{Hucker}.

Note that the behavior observed along the line ${\rm \%Sr=\%Zn}$ is in
qualitative agreement with the experimental results available for Li doped 
LCO. It is well known that the system is an insulator up to the maximum 
possible percentage of Li (50\%)\cite{Yoshi,Sarrao} 
and at 10\% doping spin IC is
not observed\cite{Bao} which was very unexpected.
Thus, these results indicate that qualitatively, 
Li doping should be equivalent to equal 
percentages of Zn and Sr doping. However,
in order to represent Li doping more accurately, it has to be
considered that Li is less positively charged than Zn and Cu and thus it 
should act as an attractor of mobile holes. To mimic this behavior the
hopping towards the impurity sites should be reduced. In our previous 
results it has been observed that t=1 tends to greatly localize the mobile
holes
since the electronic density at the impurity sites is only 0.03.  
Thus, any smaller value of t will not introduce important qualitative 
changes but, for completness, and since it has been used in other 
approaches,\cite{Neto} the results for t=0 near NMI are also presented. 
In this case the MH get trapped by the impurity and become
totally 
localized. This can be seen in Fig.4-a where the density of states is shown 
for different percentages of Li dopping. Localized states (marked in black), 
which do not disperse with momentum, proportional to
the number of added holes appear in the density of states, and the chemical
potential lies inside a gap indicating that the system is insulator as
experimentally found.\cite{Kastner,Sarrao}. 
Incommensurability is not observed as 
it can be seen in Fig.4-b where the structure factor is displayed.
A snapshot is presented in Fig.4-c.

Summarizing, the effect of adding non-magnetic impurities 
to a model of interacting fermions and 
classical spins has been investigated using MC techniques. When the percentage
of mobile holes 
is larger than the NMI, charge and spin IC is observed. The NMI act as
pinning centers for the horizontal and vertical stripes that are observed in
the absence of vacancies. The pinned stripes tend to become diagonal and 
surround AF domains. Thus, the observed incommensuration may not be identical
to the one observed in LSCO.
When the percentage of MH is equal to or smaller than the NMI, the mobile
holes are
trapped by the impurities and the system remains insulator and magnetically
commensurate. Thus, based on our study, it is predicted 
that incommensuration should not be 
experimentally observed in this situation, and that samples with equal
percentages of Sr and Zn should behave similarly to the Li doped ones.

A.M. is supported by NSF under grant DMR-9814350.
Additional support is provided by the National High Magnetic Field Lab 
and MARTECH.


\begin{references}

\bibitem{Cheong} S-W.~Cheong {\it et al.}, Phys.~Rev.~Lett.~{\bf 67}, 1791
(1991); P.~Dai {\it et al.}, Phys.~Rev.~Lett.~{\bf 80},
1738 (1998); H.A.~Mook {\it et al.}, Nature {\bf 395}, 580 (1998).

\bibitem{Tran} J.M.~Tranquada {\it et al.}, Phys.~Rev.~Lett.~{\bf 78}, 338 (1997).

\bibitem{Chakra} A.~Chakraborty {\it et al.}, 
Phys.~Rev.~B{\bf 40}, 5296 (1989);
G.~Xiao {\it et al.}, ibid. {\bf 42}, 240 (1990); K.~Uchinokura {\it et al.},
 Physica
{\bf B 205}, 234 (1995).

\bibitem{Hucker} M.~H\"ucker{\it et al.}, Phys. Rev. B{\bf 59}, R725 (1999).

\bibitem{Fisk} S.-W.~Cheong {\it et al.}, Phys. Rev. B{\bf 44}, 9739 (1991).

\bibitem{Hirota} K.~Hirota {\it et al.}, Physica{\bf B 241-243}, 817 (1998).

\bibitem{Bao} W.~Bao {\it et al.}, 
to appear in PRL. Preprint, cond-mat/9909256.

\bibitem{Charlie} C.~Buhler {\it et al.} 
Phys.~Rev.~Lett.~{\bf 84}, 2690 (2000). 

\bibitem{Pines} P.~Monthoux {\it et al.}, Phys.~Rev.~B{\bf 47}, 6069
(1993); A.~Chubukov, Phys.~Rev.~B{\bf 52}, R3840 (1995);
S.~Klee and A.~Muramatsu, Nucl. Phys. {\bf B 473}, 539 (1996).

\bibitem{Schrieffer} J.R.~Schrieffer, 
J. of Low Temp.~Phys.~{\bf 99}, 397 (1995);
B.L.~Altshuler {\it et al.}
Phys.~Rev.~B{\bf 52}, 5563
(1995).

\bibitem{Fedro} C.-X.~Chen {\it et al.}
Phys.~Rev.~B{\bf 43}, 3771 (1991).

\bibitem{yuno} E.~Dagotto {\it et al.}, Phys. Rev. {\bf B 58}, 6414 (1998).

\bibitem{Hilscher} G.~Hilscher {\it et al.}, Z.~Phys.B{\bf 72}, 461 (1988).

\bibitem{Kar} K.~Karpinska {\it et al.}, preprint cond-mat/9911149.

\bibitem{Fuku} Y.~Fukuzumi {\it et al.}, Phys. Rev. Lett. {\bf76}, 684 (1996).

\bibitem{yamada}K.~Yamada {\it et al.}, Phys.~Rev.~B{\bf 57}, 6165 (1998).

\bibitem{YBC} W.A.~MacFarlane {\it et al.}, preprint, cond-mat/9912165; Y.~Sidis
{\it et al.}, preprint, cond-mat/9912214; K.~Segawa and Y.~Ando, Phys.~Rev.~B
{\bf 59}, R3948 (1999); H.F.~Fong {\it et al.}, Phys. Rev. Lett. {\bf82}, 1939 
(1999).

\bibitem{Tran2} J.M.~Tranquada {\it et al.}, Phys.~Rev.~{\bf B 54}, 7489 
(1996).

\bibitem{com} Since PBC are used the $\pi-$shift is not observed when it 
would cause spin frustration.\cite{Charlie}

\bibitem{e2} G.~Martins {\it et al.}, preprint.

\bibitem{endoh} T.~Sato {\it et al.}, preprint, to appear in Phys. Rev. Lett.

\bibitem{Neto} A.H.~Castro Neto and A.V.~Balatsky, cond-mat/9805273;
G.~Martins {\it et al.}, Phys. Rev. Lett. {\bf78}, 3563 (1997).

\bibitem{Yoshi} Y.~Yoshinari {\it et al.}, Phys. Rev. Lett. {\bf77}, 2069 (1996).

\bibitem{Sarrao} B.J.~Suh {\it et al.}, Phys. Rev. Lett. {\bf81}, 2791 (1998).

\bibitem{Kastner} M.~Kastner {\it et al.}, Phys.~Rev.~B{\bf 37}, 111 (1988);
J.~Sarrao {\it et al.}, Phys.~Rev.~B{\bf 54}, 12014 (1996).


\end{references}
\end{document}